# Deep-Learning Inversion Method for the Interpretation of Noisy Logging-While-Drilling Resistivity Measurements

Kyubo Noh, David Pardo, and Carlos Torres-Verdín, *Member, IEEE*

*Abstract*—Deep Learning (DL) inversion is a promising method for real-time interpretation of logging-while-drilling (LWD) resistivity measurements for well navigation applications. In this context, measurement noise may significantly affect inversion results. Existing publications examining the effects of measurement noise on DL inversion results are scarce. We develop a method to generate training data sets and construct DL architectures that enhance the robustness of DL inversion methods in the presence of noisy LWD resistivity measurements. We use two synthetic resistivity models to test three approaches that explicitly consider the presence of noise: (1) adding noise to the measurements in the training set, (2) augmenting the training set by replicating it and adding varying noise realizations, and (3) adding a noise layer in the DL architecture. Numerical results confirm that the three approaches produce a denoising effect, yielding better inversion results—in both predicted earth model and measurements—compared not only to the basic DL inversion but also to traditional gradient-based inversion results. A combination of the second and third approaches delivers the best results. The proposed methods can be readily generalized to multi-dimensional DL inversion.

## I. INTRODUCTION

AS the demand for the drilling of complex geological structures and production-efficient well placement increases, well navigation, or well geosteering, which dynamically adjusts the well trajectory, has become an essential technique. Among the various geophysical measurements acquired during drilling, logging-while-drilling (LWD) resistivity measurements are pivotal for well geosteering because resistivity contains comprehensive information about the spatial distribution of surrounding rocks and their saturating fluids [1]-[2]. Thus, real-time processing and interpretation of LWD resistivity measurements has become an important tool for active well geosteering decisions.

The interpretation of geophysical measurements is most often performed via inversion techniques. Conventional inversion employs deterministic or stochastic methods. With the recent widespread utilization of machine learning (ML) algorithms, inversion using ML—and in particular, deep learning (DL)—has become a valuable alternative to traditional inversion methods [3]-[5]. For example, [6] inverts 1.5-dimensional (1.5D) LWD resistivity measurements with a deep neural network (DNN) and a rigorous forward model by invoking a loss function that combines both model- and data-misfit; in [7], authors construct a DNN approximation of both forward and inverse problems; in [8], authors combine a gradient-based and a ML inversion method to enhance the generalization of training conditions; [9] considers multiple DL inversion architectures designed for varying geological situations for a 2.5-dimensional (2.5D) inversion of possibly fractured rocks.

Under real drilling conditions, LWD resistivity measurements are inevitably contaminated by noise. This adverse condition affects the stability and reliability of the inversion. Thus, noise-robustness becomes a key feature of an inversion algorithm. Among the traditional inversion methods, stochastic approaches are naturally designed to evaluate the uncertainty of inversion results in the presence of noisy and inadequate measurements [10]. In gradient-based methods, alternative approaches have been used, such as error bars to evaluate uncertainty and non-uniqueness—including noise effects [11]—and Huber-norm-based inversion to minimize the effect of Gaussian and non-Gaussian noise [12].

However, documented studies examining the robust implementation of DL inversion for noisy LWD resistivity measurements are scarce. A few pioneering works in this direction show that training with noisy measurements enhances the robustness of the inversion [6], [8]. Adding noise in the training data set is the most common approach used to enhance the generalization capacity of a DL method [13]; it has been shown that adding noise during training functions as a traditional Tikhonov regularization [14]. Despite these pioneering works, further studies are needed concerning the

This work reported in this paper was supported by the University of Texas at Austin's Research Consortium on Formation Evaluation, jointly sponsored by Aramco, BHP Billiton, BP, Chevron, CNOOC International, ConocoPhillips, COSL, ENI, Equinor ASA, Haliburton, INPEX, Lundin-Norway, Oil Search Alaska, Oxy, Petrobras, Repsol, Schlumberger, Todd Energy, Total, and Wintershall. The authors also acknowledge the Texas Advanced Computing Center (TACC) at the University of Texas at Austin for providing HPC resources that have contributed to the research results reported in this paper. URL: http://www.tacc.utexas.edu

K. Noh and C. Torres-Verdín are with The University of Texas at Austin, Austin, TX 78712 (email: kyubonoh@gmail.com, cverdin@austin.texas.edu). D. Pardo is with University of the Basque Country (UPV/EHU), Leioa, Spain; Basque Center for Applied Mathematics (BCAM), Bilbao, Spain; and Ikerbasque, Bilbao, Spain. (email: dzubiaur@gmail.com).

adverse effects of noise on DL inversion results and possible effective remedies.

We study the effect of noise on the DL inversion of LWD resistivity measurements and propose three strategies to mitigate the effect of noise on inversion results, namely: (1) to add synthetic noise in the training data set; (2) to incorporate synthetic noise using augmented training data; and (3) to include a layer that adds noise in the DL architecture. To numerically assess the performance of these alternatives, we employ a 1.5D DL inversion method to compare the results of the aforementioned methods with those obtained from conventional gradient-based inversion. The comparison includes a condition where different noise levels are employed for the training and test data sets.

This paper is organized as follows: In Section II, we describe our DL inversion problem, including the DL loss function and the DL architectures we employ to perform the noise-robust inversion. Section III details our well-trajectory parameterization, assumed measurement system, noise contamination, and the construction of synthetic data sets for DL training. In Section IV, we employ two synthetic resistivity models to evaluate the effect of our three strategies in the inversion of noisy measurements. Section V summarizes the main conclusions

## II. DEEP LEARNING INVERSION FOR RECONSTRUCTING RESISTIVITY MODELS

### A. Objective (Loss) Function

To train DL architectures that perform inversion of subsurface models from the measurements, we solve the following minimization problem:

$$\left(F_{\phi^*}, I_{\theta^*}\right) := \arg\min_{\phi \in \Phi, \theta \in \Theta} \left\{ \left\|\left(F_\phi \circ I_\theta\right)(\mathbf{M}, \mathbf{T}) - \mathbf{M}\right\|_2^2 + \left\|F_\phi(\mathbf{P}, \mathbf{T}) - \mathbf{M}\right\|_2^2 \right\}, \quad (1)$$

where $\mathbf{P}$, $\mathbf{M}$, $\mathbf{T}$ denote subsurface properties, measurements, and well trajectory, respectively, and $F_\varphi$ and $I_\theta$ are the DL approximations of the forward and inverse functions, respectively. Symbols $\varphi$ and $\theta$ represent sets of trainable weights in the DL approximations. In Eq. (1), the second additive component measures the forward approximation misfit; the first one ensures that the composition of the inverse with the forward operator approximates the identity, and it can be interpreted as an encoder-decoder DL architecture. The above loss function produces the operators $F_{\varphi^*}$ and $I_{\theta^*}$. The above objective function properly deals with the possible non-uniqueness of the inverse solution; see [7] for details.

### B. Deep Learning Architecture

For solving Eq. (1), we use a deep residual neural network (ResNet) [15]-[16] with convolutional layers [17]. Our deep ResNet consists of several residual blocks, which include convolutional, batch normalization, and activation layers. The identity mapping between each residual block enables a successful training of the DL architecture by preventing vanishing gradients and degradation in deep architectures [16].

In DL architectures, we can apply noise to the hidden layers to act as a regularizer and enhance robustness against noise [13]. This procedure can be implemented in various ways. Work [18] shows that adding noise to the gradient at every training step avoids data overfitting and increases the training performance of deep architectures. In [19], the authors propose to use a noisy activation function to enhance randomness in the hidden state. Reference [20] proposes to inject noise in the hidden states of the autoencoders—referred to as noisy autoencoders. Following the latter results, we add noise to our DL architectures for the inversion of noisy borehole resistivity measurements. Among the existing approaches to add noise in the DL architecture, we follow the one introduced in [20].

We build two types of DL architectures: a noise-free ResNet and a ResNet with a hidden noisy layer. Next, we compare the DL inversion results obtained with both architectures. The vanilla form follows the general type of ResNet constituted by residual blocks with batch normalization, ReLU [21] activation, 1D convolution, and Dropout layers [22]. For the second architecture, to enhance the noise components in the hidden states, we add a Gaussian noise component after every convolutional layer. We select two values of the standard deviation of the Gaussian noise to control the intensity of the hidden noisy layer. Consequently, we have three DL architectures: one noise-free ResNet and two noisy ResNets detailed in Table 2. Using the selected ResNet architectures, we define both forward and inverse approximators ($F_\phi$ and $I_\theta$ in (1)) by setting adequate input and output shapes for each approximator. The Dropout layers also act as strong regularizers by functioning as noise in hidden states [13]. Appendix A provides further details about the DL architectures employed in this study.

## III. DATA PREPARATION AND TRAINING

In what follows, we describe the subsurface and well trajectory parameterizations, measurement system, and measurement noise employed to generate the DL training and validation data sets.

### A. Subsurface Parameterization

We consider a piecewise-1D inversion approach, which is commonly used in geosteering applications [23]. Additionally, the Earth model is assumed to be a three-layer medium, as shown in Fig. 1. The host layer is anisotropic with horizontal and vertical resistivities denoted as $\rho_h$ and $\rho_v$, respectively; the upper and lower layers are isotropic with resistivities $\rho_u$ and $\rho_l$, respectively. We also consider the vertical distance to the upper and lower layers—denoted as $d_u$ and $d_l$, respectively—and the layer dip angle, denoted as $\beta$. According to the physics of our problem, we restrict the resistivity values to the range [0.1, 1000] ohm-m, the anisotropy factor $\rho_v/\rho_h$ to the range [1, 10], the distance values $d_u$ and $d_l$ to the range [0.01, 5] m, and the dip angle $\beta$ to the range [-30, 30] degrees. Furthermore, we sample resistivity and distance values in a log-scale, and the dip angle in linear scale.

### B. Well Trajectory Parameterization

For our geosteering application, we consider high-angle well trajectories with a dip angle within the range [80, 100] degrees. We reconstruct each three-layer model from a 2 m-long trajectory segment (with a 0.1-meter sampling interval) given by 21 measurement points. Additionally, a maximum of 0.3-degree deviation of the drilling trajectory is assumed in a 2 meters section; hence, the initial dip angle is within the range [80.3, 99.7] while well deviation from one measurement point to the next is within the range [-0.03, 0.03] degrees.

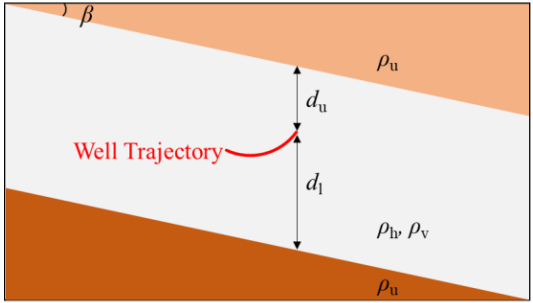

**Figure 1.** Graphical representation of the one-dimensional resistivity model assumed in this study. The red line indicates the local well trajectory.

*C. Measurement System and Measurement Noise*

We consider two logging instruments: a short-sensing LWD instrument and a deep-sensing instrument—see Fig. 2. The short LWD instrument includes five pairs of transmitters equally spaced from the tool center and operates at 2 MHz; on the other hand, the deep instrument only includes one transmitter-receiver pair separated by 12 meters and operating at 10 kHz. From the five transmitter-receiver pairs of the short LWD instrument and the transmitter-receiver pair of the deep instrument, we calculate attenuation and phase components of one co-axial, two co-planar, and one cross-coupling components (denoted as *Geosignal* component). The Appendix in [7] describes in detail the corresponding postprocessing implemented to the acquired measurements to replicate commercial implementations of similar systems. In summary, for each model, the input data consist of 6×4 types of measurements, each one having a length equal to 21. For the data sets selected in the numerical experiments, we introduce noise in both training and test data sets. We implement a Gaussian distribution of noise with two intensities—see Table 1.

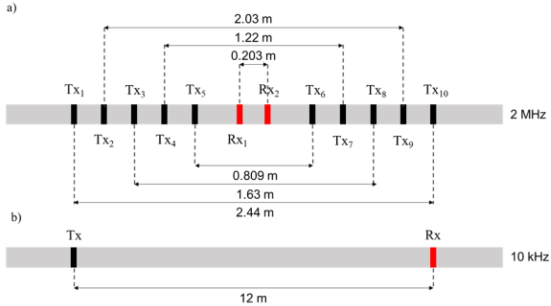

**Figure 2**. a) Conventional LWD and b) Deep azimuthal logging instruments assumed in this study. Each transmitter and receiver include three mutually orthogonal coils.

*D. Training Data Sets and Training Process*

We generate four training data sets with varying noise types for our comparative study about noise effects on inversion results. The first data set contains 100K samples. We use a semi-analytic numerical method to perform the 1.5D simulations (see [23] and references therein), where the computation takes approximately 20 CPU hours using a computer equipped with Intel(R) Core i7-8700 CPU. Using this first data set, we generate a second and a third data set by adding weak and strong noise, respectively. The fourth data set contains 300K samples. To generate the fourth data set, we first triplicate the first data set, then we contaminate it with varying weak noise. With this fourth data set, the inversion module is trained to infer identical resistivity models from each set of three measurement samples contaminated with different noise values. This fourth data set will be referred to as an augmented weakly noisy data set.

We train each DL architecture independently for each data set. In so doing, we select 80% of the samples for training, while the remaining 20% are equally divided between validation and test data sets. The training process for each of the first three data sets takes approximately 8 hours using an Nvidia GTX machine equipped with Intel Xeon E5-2620 v4 CPU and four Nvidia 1080-TI GPUs, while that for the fourth data set takes 10 hours with a given set of hyperparameters.

**Table 1.** Condition for measurement noise in training and validation data sets. We consider zero-mean Gaussian noise with the maximum intensities given below.

|  | LWD measurement [Att., Phase shift] | Deep measurement [Att., Phase shift] |
|---|---|---|
| Weak noise | [0.1 dB, 0.4 degrees] | [0.004 dB, 0.4 degrees] |
| Strong noise | [0.5 dB, 2.0 degrees] | [0.02 dB, 2.0 degrees] |

## IV. NUMERICAL RESULTS

To assess the effect of noise contamination on DL inversion results and propose possible remedies, we utilize two synthetic well geosteering problems: a simple three-layer isotropic model and a complex three-dimensional anisotropic model. For the assessment, we compare the inversion results by using varying combinations of noise in the (a) measurements (see Table 1), (b) training data (see Table 2), and (c) the DL architecture (see Table 3).

*A. Simple Three Layer Isotropic Model*

We consider the simple three-layer earth model described in Fig. 3a. The model consists of an isotropic conductive layer of 1 Ohm-m within an isotropic resistive background—100 Ohm-m, where the well trajectory starts from the resistive layer and penetrates the conductive layer.

**Table 2.** Different deep learning architectures according to the noise intensity added to the architecture.

| Architecture | Standard Deviation |
|---|---|
| DL—Noise Free | - |
| DL—Noise Weak | $10^{-5}$ |
| DL—Noise Strong | $10^{-1}$ |

**Table 3.** Different training data sets with varying measurement noise level and number of samples. The level of noise intensity is shown in Table 1.

| Dataset | # of Samples |
|---|---|
| Training—Noise Free | 100K |
| Training—Noise Weak | 100K |
| Training—Noise Weak Augmented | 300K |
| Training—Noise Strong | 100K |

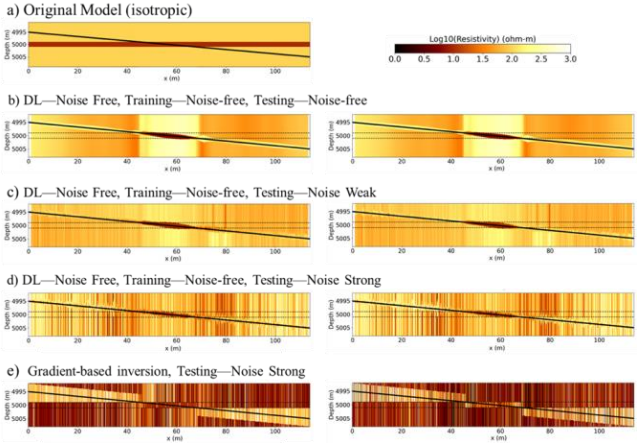

**Figure 3.** (a) Original isotropic resistivity model and resistivity models estimated with (b) deep learning inversion of noise-free data, (c) deep learning inversion of weakly noisy data, (d) deep learning inversion of strongly noisy data, and (e) gradient-based inversion of strongly noisy data. Left- and right-hand columns of inversion results show horizontal and vertical resistivities, respectively.

*1) Vanilla DL inversion of noisy measurements*

Fig. 3b shows the DL inversion results obtained with noise-free measurements using a noise-free DL architecture trained with a noise-free data set—this situation corresponds to experiment 1 in Table 4. We observe an adequate reconstruction of formation resistivity: the inversion correctly detects the conductive layer embedded within the resistive host formation; it also detects the bed boundaries nearby the well trajectory (e.g., at horizontal location *x* within the ranges [35, 40], [50, 55], [60, 65], and [75, 80] m). However, there are some incorrect estimations of bed boundaries located far away from the well trajectory. This occurs due to the limited depth of investigation of the assumed logging instruments.

Fig. 3c and 3d show DL inversion results obtained when using the same architecture as in Fig. 3b, but with the weakly and strongly noisy test data, respectively: they correspond to experiments 2 and 3 in Table 4, respectively. We observe a rough approximation of the original formation, where the prediction of layer boundaries is inadequate. The formation model mismatch is greater with the strongly noisy measurements (Fig. 3d) than with the weakly noisy measurements (Fig. 3c). As a reference to the DL inversion results obtained for the case of strongly noisy measurements in Fig. 3d, we display in Fig. 3e the gradient-based 1.5D inversion results obtained with the strongly noisy measurements, which corresponds to experiment 4 in Table 4. The deterministic inversion includes only six unknowns (distances are fixed), because otherwise bed-boundaries would be significantly misplaced. Inversion results lack spatial continuity and exhibit significant artifacts, especially concerning vertical resistivities.

**Table 4.** Root-mean-square error (RMSE) and difference (RMSD) of inversion results corresponding to the simple three-layer model used in Fig. 3, Fig. 5, and Fig.6. The minimum RMSE and RMSD for the strongly noisy input measurement are shown with bold numbers.

| Ex. # | Inversion Method | RMSE (Att. / Pha. shift) | RMSD (Att. / Pha. shift) |
|---|---|---|---|
| 1 | • DL—Noise Free<br>• Training—Noise Free<br>• Testing—Noise Free | (0.074, 0.469) | (0.074, 0.469) |
| 2 | • DL—Noise Free<br>• Training—Noise Free<br>• Testing—Noise Weak | (0.125, 0.716) | (0.092, 0.516) |
| 3 | • DL—Noise Free<br>• Training—Noise Free<br>• Testing—Noise Strong | (0.334, 1.377) | (0.194, 0.713) |
| 4 | • Gradient-based<br>• Testing—Noise Strong | (0.346, 1.655) | (0.282, 1.194) |
| 5 | • DL—Noise Free<br>• Training—Noise Strong<br>• Testing—Noise Strong | (0.290, 1.370) | (0.144, 0.530) |
| 6 | • DL—Noise Free<br>• Training—Noise Weak<br>• Testing—Noise Strong | (0.314, 1.310) | (0.185, 0.706) |
| 7 | • DL—Noise Free<br>• Training—Noise Weak Augmented<br>• Testing—Noise Strong | (0.302, 1.274) | (0.125, 0.502) |
| 8 | • DL—Noise Strong<br>• Training—Noise Weak<br>• Testing—Noise Strong | (0.315, 1.328) | (0.152, 0.626) |
| 9 | • DL—Noise Weak<br>• Training—Noise Weak<br>• Testing—Noise Strong | (0.291, 1.257) | (0.093, 0.499) |
| 10 | • DL—Noise Weak<br>• Training—Noise Weak Augmented<br>• Testing—Noise Strong | **(0.285, 1.219)** | **(0.084, 0.483)** |

Figs. 4a and 4b compare the original noise-free measurements to those with strong noise, as well as the predictions from the inversion results obtained using DL and gradient-based approaches—they correspond to experiments 3 and 4 in Table 4, respectively. Among the existing 24 components, we display the co-axial component of the shortest transmitter-receiver pair included in the LWD instrument.

Although both methods reconstruct to some extent the original measurements, gradient-based inversion results exhibit larger differences with respect to those obtained with the noise-free original measurements. To quantitatively evaluate the performance of data reconstruction for the inversion result, we display in Table 4 the root-mean-squared error (RMSE) between the noise-contaminated measurements and the numerically simulated measurements. We also show the root-mean-squared difference (RMSD) between the original noise-free and numerically simulated measurements. All 24 components are used to compute the RMSE and RMSD values. Comparison of the results obtained from experiments 1 to 3 indicates that both RMSE and RMSD increase as measurement noise increases. We also observe lower RMSD than RMSE values for all experiments that indicates reconstruction capability of original noise-free measurements of DL inversion. Gradient-based inversion results also exhibit lower RMSD than RMSE, but the difference is smaller than that of the DL inversion results. In summary, the gradient-based inversion is less robust to measurement noise than the vanilla DL inversion.

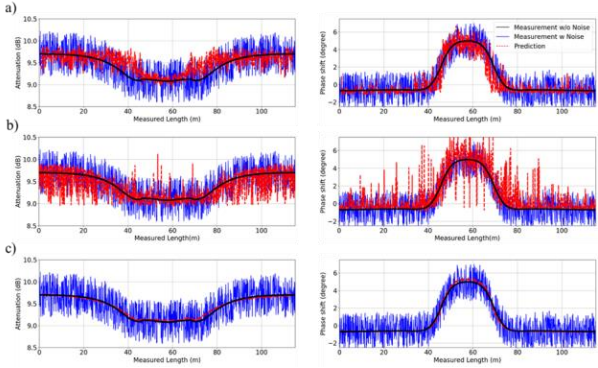

**Figure 4.** Comparison of noise-free measurements (black solid lines), strongly noise-contaminated measurements (blue solid lines), and predicted measurements obtained with (a) deep learning inversion without measurement noise during training, (b) gradient-based inversion, and (c) deep learning inversion with measurement noise during training. Left- and right-hand columns display attenuation and phase difference components, respectively, of nearest Tx-Rx pair corresponding to the conventional LWD tool shown in Fig. 2a.

*2) Adding measurement noise of a single distribution*

In this subsection, we independently train the DL architectures by using training data sets with measurement noise: "DL—Noise Strong" and "DL—Noise Weak" in Table 3 and then apply the DL architectures to the test data with strong measurement noise. These conditions correspond to experiments 5 and 6 in Table 4, respectively. Fig. 5a—trained with strong noise—shows a more spatially continuous and accurate reconstruction of the distribution of electrical resistivity than Fig. 3d—trained with noise-free data. Thus, inversion results become more reliable as the well trajectory penetrates the conductive layer embedded in a resistive host. Fig. 5b is obtained by applying a DL architecture trained with a weaker level of measurement noise than the test data set. Thus, this case resembles real-world operations where the actual noise level is uncertain. Compared to the noise-free training data set—Fig. 3d—, Fig. 5b shows a more spatially continuous resistivity distribution. Training with a different noise level than the actual measurement noise level seems helpful to delineate the actual resistivity structure. Table 4 shows the RMSE and RMSD values associated with the inversion results obtained for experiments 5 and 6. RMSE and RMSD values of experiment 5 are lower than those of experiment 6, which indicate that a better resistivity reconstruction occurs when noise levels in the measurements and training data sets are close to each other. However, even when the noise levels are different, we also observe a better reconstruction of the resistivity distribution when including noise in the training data set than when employing noise-free training data—compare results of experiments 6 and 3.

*3) Adding measurement noise of two different distributions*

*Numerical Experiments are under process*

*4) Training data set with augmented measurement noise*

We now use a training data set with an augmented measurement noise, which corresponds to the "Training—Noise Weak Augmented" described in Table 3. We invert strongly noisy measurements, which corresponds to experiment 7 in Table 4. Fig. 5c shows the inversion result. Improved inversion results are obtained compared to the result obtained without noise augmentation in Fig. 5b. In particular, we adequately estimate the bed boundaries at $x$ within the range [45, 50] compared to the result shown in Fig. 5b. In addition, the upper and lower bed boundaries become more clearly delineated when the well trajectory is positioned within the conductive layer. Comparison of the RMSE and RMSD obtained with and without noise augmentation (experiments 6 and 7 in Table 4, respectively), indicates that using the noise-augmentation approach better reconstructs both noise-free and noisy measurements.

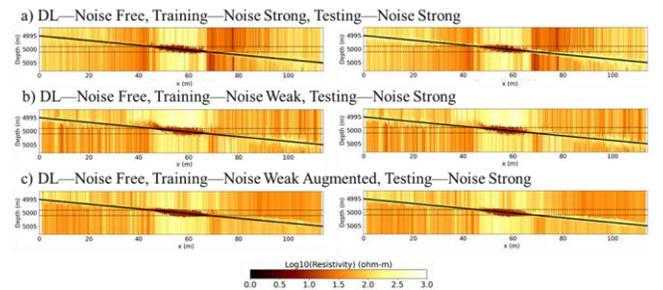

**Figure 5.** Inverted resistivity models predicted by the deep learning inversion algorithms of strongly noisy data. We consider three deep learning architectures trained with: (a) strong measurement noise, (b) weak measurement noise, and (c) augmented weak measurement noise. Left- and right-hand columns show inverted horizontal and vertical resistivities, respectively.

*5) Adding a noise layer to the DL architecture*

To enhance the noise robustness of the DL inversion method, we train two DL architectures with a weakly noisy training data set by varying the standard deviation for Gaussian architecture noise: "DL—Noise Strong" and "DL—Noise Weak" in Table 2. Experiments 8 and 9 in Table 4 are conducted by inverting strongly noisy measurements; inversion results are shown in Fig. 6a and 6b, respectively. Fig. 6a (obtained with DL—Noise Strong) shows a spatially smooth resistivity distribution in the horizontal direction and a better reconstruction of the layered structure compared to the inversion result obtained without either measurement noise in the training data or the architecture noise shown in Fig. 3d. Close to the center of the model, inversion results adequately predict a conductive layer embedded within a resistive background. However, we observe inaccurate estimations in other regions in terms of bed boundary locations and horizontal resistivity near layer boundaries. The relatively high architecture noise causes the DL architecture to focus on the general trend rather than on small features. Fig. 6b shows the inversion result corresponding to the strongly noisy test data obtained with the "DL—Noise Weak". It exhibits better estimations of bed boundary locations (without losing the resolving power on the three-layered structure) compared to the inversion result obtained with the strongly noise DL architecture in Fig. 6a. After applying weak architecture noise, inversion results clearly show both bed boundaries and a three-layer structure, without the strong artifact observed in Fig. 5b. The effect of architecture noise can also be observed in RMSE and RMSD values in Table 4. The result obtained with weak architecture noise exhibits lower RMSE and RMSD values than those corresponding to the strong architecture noise. Important differences between the two architectures indicate that adding architecture noise can only be effective when that the magnitude of the architecture noise is carefully tuned [18].

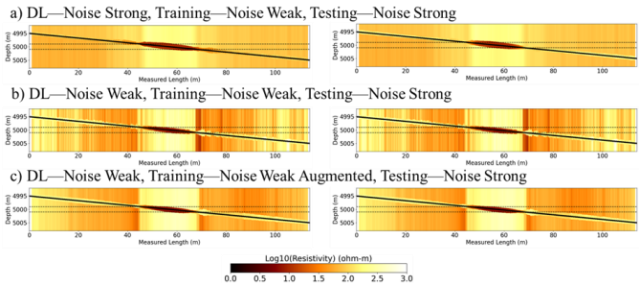

**Figure 6.** Inverted resistivity models obtained with deep learning inversion algorithms of strongly noisy data. We consider three deep learning architectures trained with: (a) noise-free measurements and strong architecture noise, (b) noise-free measurements and weak architecture noise, and (c) augmented weak measurement noise and optimal architecture noise. Left- and right-hand columns show inverted horizontal and vertical resistivities, respectively.

*6) Combined augmented measurement noise and architecture noise*

We combine both augmented measurement noise in the training data set and architecture noise: "Training—Noise Weak Augmented" and "DL—Noise Weak" in Tables 3 and 2, respectively. Fig. 6c displays the corresponding inversion results obtained with the strongly noisy test data, which corresponds to experiment 10 in Table 4. The inversion result predicts a resistivity distribution similar to that obtained without the augmenting approach in Fig. 6b, but with a smoother resistivity distribution in the horizontal direction.

Results shown in Fig. 6c better reconstruct the three-layer structure at the center of the model and bed boundaries compared to the result in Fig. 3d obtained without any treatment for measurement noise. Fig. 6c shows a similar resistivity distribution to that obtained with noise-free measurements; the corresponding predicted measurements (shown in Fig. 4c) also indicate a good reconstruction of the original noise-free measurements. RMSE and RMSD values associated with Fig. 6c and shown in Table 4 indicate the lowest performance among the results obtained with strong measurement noise. Specifically, the RMSD result shows the high capacity of the inversion to predicting denoised original measurements when employing both augmented measurement noise and architecture noise.

*B. Three-Dimensional Anisotropic Model*

We now apply our DL inversion approach to a complex 3D model. Fig. 7 shows a 3D view of the model along with a 3D well trajectory. We consider a model consisting of ten layers with varying dip angles and two vertical fault planes. Fig. 8 shows plane views of horizontal and vertical resistivity across the center of the model in the *y*-direction with the 3D well trajectory projected in that plane. Fig. 9a shows the horizontal and a vertical resistivity distribution of the corresponding three-layer model along the 3D well trajectory. We simulate 1D resistivity measurements using the three-layer model and contaminate the simulations using strong noise.

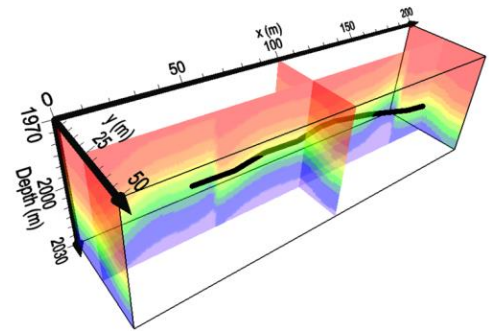

**Figure 7.** Three-dimensional plot of a complex earth model consisting of ten layers and two vertical faults located at *x*=50 and *x*=170 m. The black solid line identifies the well trajectory while the background color indicates the layer indexes.

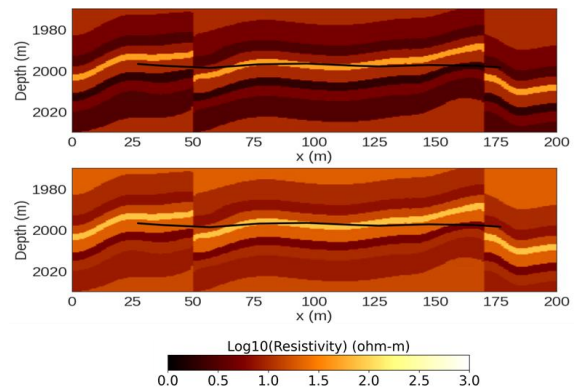

**Figure 8.** Cross-sections of (a) horizontal and (b) vertical resistivity model in *x-depth* space at *y*=25 m shown in Fig.7. The three-dimensional

well trajectory is projected onto the two-dimensional space and displayed with a black line.

Figs. 9b, 9c, and 9d compare inversion results obtained with: (a) DL without any noise treatment—experiment C-1 in Table 5—, (b) DL inversion with augmented weak noise for training and weak architecture noise—experiment C-2 in Table 5—, and (c) gradient-based 1D inversion—experiment C-3 in Table 5. The DL inversion result obtained without noise treatment in Fig. 9b only shows reliable reconstruction of the host resistivity values along the well trajectory. However, it fails to properly estimate a the three-layer structure and the bed boundaries. Fig. 9c, obtained after applying augmented measurement noise and optimal architecture noise, shows a spatially smooth resistivity distribution in the horizontal direction. Even though some inaccuracies appear in the prediction of bed boundaries, when the latter are close to the well trajectory (e.g., at $x$ values of approximately 120, 140, and 160 m), inversion results estimate the actual location of the bed boundaries. Despite the fact that the gradient-based inversion result shown in Fig. 9d was obtained with the exact bed boundaries, results lack spatial continuity in the horizontal direction and the estimated host resistivities are degraded compared to those obtained with the DL inversion.

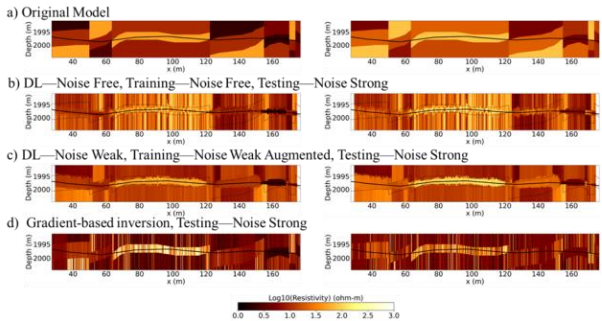

**Figure 9.** (a) Original three-layer resistivity model along the well trajectory. Panels b), c) and d) display inverted resistivity models for the strongly noisy data obtained with (b) deep learning inversion trained without measurement and architecture noise, (c) deep learning inversion trained with augmented weak measurement noise and optimal architecture noise, and (d) gradient-based inversion. Left- and right-hand columns show inverted horizontal and vertical resistivities, respectively.

**Table 5.** Root-mean-square error (RMSE) and difference (RMSD) of inversion results corresponding to the inverted model shown in Fig. 9.

| Ex. # | Inversion Method | RMSE (Att. / Pha. shift) | RMSD (Att. / Pha. s |
|---|---|---|---|
| C-1 | • DL—Noise Free<br>• Training—Noise Free<br>• Testing—Noise Strong | (0.442, 2.256) | (0.336, 1.6 |
| C-2 | • DL—Noise Weak<br>• Training—Noise Weak Augmented<br>• Testing—Noise Strong | (0.377, 1.817) | (0.186, 0.9 |
| C-3 | • Gradient-based<br>• Testing—Noise Strong | (0.346, 1.655) | (0.282, 1.1 |

Comparisons of the original noise-free, observed noisy, and predicted measurements obtained from the inversion results shown in Figs. 9b, 9c, and 9d are shown in Figs. 10a, 10b, and 10c, respectively. We reproduce the original noise-free measurements of the DL inversion with augmented training data noise and architecture noise. Table 5 compares the RMSE and RMSD values of the inversion results shown in Fig. 9. As in the previous numerical model, the DL inversion result with augmented training noise and architecture noise exhibits lower RMSE and RMSD values than those associated with the gradient-based and basic DL inversion results. The improvement in RMSD is remarkable in the DL inversion results when using a noisy training data set and architecture noise. After applying the augmented measurement training noise and a proper level of the architecture noise, the DL inversion delivers reliable results.

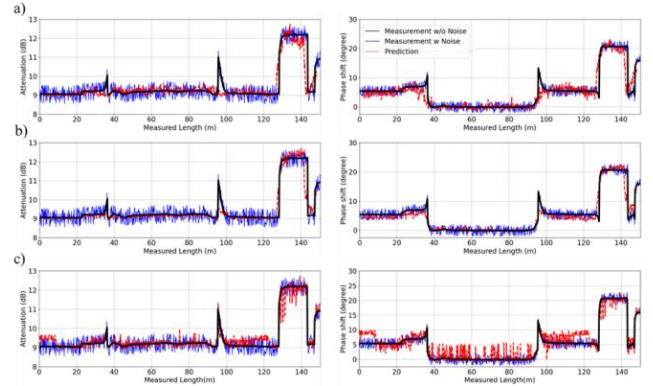

**Figure 10.** Comparison of noise-free measurements (black solid lines), strongly noise-contaminated measurements (blue solid lines), and predicted measurements obtained with (a) deep learning inversion trained without measurement and architecture noise, (b) deep learning inversion with augmented weakly noisy data and optimal architecture noise, and (c) gradient-based inversion. The corresponding inversion results are shown in Fig. 9b, Fig. 9c, and Fig. 9d, respectively. Left- and right-hand columns display attenuation and phase difference components, respectively, of the nearest Tx-Rx pair corresponding to the conventional LWD tool shown in Fig. 2a.

## V. CONCLUSIONS

We investigated the effect of measurement noise in the DL inversion of borehole resistivity measurements using two synthetic models. We then proposed various remedies to improve the inversion results; they consist of training DL networks using a noise-contaminated—and possibly augmented—training data set and adding a noise layer within the architecture. Results show that both gradient-based and DL inversion trained without considering noise are vulnerable to noisy measurements. Adding noise to the training data set enhances the noise robustness of the DL inversion in both model and data spaces. The effectiveness of the noise contamination strategy in the measurements increases when selecting the same noise level as the one existing in the actual input measurements. Because the exact noise level for a specific acquisition environment is unknown, we propose to use augmented noisy training data sets with different noise levels to improve inversion performance. Adding a noise layer in the DL architecture enhances the generalization capacity and noise robustness of the DL inversion. We use a grid-searching strategy to derive the proper level of architecture noise. The combination of augmented measurement noise with a noise layer in the DL architecture provides the best inversion performance for the noisy test data in both the predicted earth model and predicted measurements. Inversion methods developed in this paper are dimension independent. Therefore, we expect the proposed remedies for the DL inversion of noisy measurements to remain valid for multi-dimensional inversion cases.

APPENDIX

*A. Neural Network Architecture*

For the approximation of forward and inverse functions, we employ the neural network architectures described in Table A1. We consider the same structure of ResNet for both forward and inverse modules with proper input and output shapes for each purpose. We also consider fully connected layers for the forward module. The fully connected layers with ten series of dense, batch normalization, ReLU, and Dropout layers are implemented as a feature extractor to properly feed to the ResNet the input resistivity structure. Following the method in the original work on ResNet [15], down-sampling is performed at the beginning of the groups of residual blocks starting from the second block (i.e., Conv 2_1, Conv 3_1, Conv 4_1, Conv 5_1, Conv 6_1 in Table A1) with a stride of two. The use of Gaussian noise layers after every 1D convolutional layer in the ResNet architecture is intended to appraise the effectiveness of the approach and its robustness in the presence of measurement noise.

**Table A1.** Description of our neural network architecture. We consider independent ResNet architectures for the forward and inverse modules. Numbers inside curly, angle, and square brackets denote the shape of vectors or matrices, number of nodes for dense layers, and the shape and numbers of convolution filters, respectively. The possible inclusion of a Gaussian noise layer after the 1D convolutional layers is described in Table 2.

| Layers | Forward | | Inverse | | Sub layers |
|---|---|---|---|---|---|
| | Model | Trajectory | Measurement | Trajectory | |
| Input | {7} | {21} | {21×24} | {21} | - |
| FC 1 | ⟨21⟩×10 | - | - | - | Dense, BN, ReLU, Dropout |
| Concatenation | {21×1}, {21} → {21×2} | | {21×24}, {21} → {21×25} | | - |
| Conv 1_x | $\begin{bmatrix}1, 32\\3, 32\\1, 64\end{bmatrix}\times 2$ | | | | 1D Conv., (Gaussian), BN, ReLU |
| Conv 2_x | $\begin{bmatrix}1, 64\\3, 64\\1, 128\end{bmatrix}\times 3$ | | | | 1D Conv., (Gaussian), BN, ReLU |
| Conv 3_x | $\begin{bmatrix}1, 64\\3, 64\\1, 128\end{bmatrix}\times 4$ | | | | 1D Conv., (Gaussian), BN, ReLU |
| Conv 4_x | $\begin{bmatrix}1, 128\\3, 128\\1, 256\end{bmatrix}\times 4$ | | | | 1D Conv., (Gaussian), BN, ReLU |
| Conv 5_x | $\begin{bmatrix}1, 128\\3, 128\\1, 256\end{bmatrix}\times 4$ | | | | 1D Conv., (Gaussian), BN, ReLU |
| Conv 6_x | $\begin{bmatrix}1, 128\\3, 128\\1, 256\end{bmatrix}\times 4$ | | | | 1D Conv., (Gaussian), BN, ReLU |
| FC 2 | ⟨462⟩ | | ⟨7⟩ | | Dense |
| Output | Measurement {21×24} | | Resistivity Model {7} | | - |